\documentclass{aastex}
\usepackage{spr-astr-addons}
\usepackage{url}

\RequirePackage{color}

\begin{document}

\title{Ion acoustic solitary structures in a collisionless unmagnetized plasma consisting of nonthermal electrons and isothermal positrons}
\slugcomment{Not to appear in Nonlearned J., 45.}
\shorttitle{IA solitary structures with nonthermal electrons and isothermal positrons}
\shortauthors{A. Paul \& A. Bandyopadhyay}

\author{Ashesh Paul\altaffilmark{1}} 
\and
\author{Anup Bandyopadhyay\altaffilmark{1}}
\affil{Correspondence: abandyopadhyay1965@gmail.com}

\altaffiltext{1}{Department of Mathematics, Jadavpur University, Kolkata
- 700032, India.}

\begin{abstract}
Employing the Sagdeev pseudo-potential technique the ion acoustic solitary structures have been investigated in an unmagnetized collisionless plasma consisting of adiabatic warm ions, nonthermal electrons and isothermal positrons. The qualitatively different compositional parameter spaces clearly indicate the existence domains of solitons and double layers with respect to any parameter of the present plasma system. The present system supports the negative potential double layer which always restricts the occurrence of negative potential solitons. The system also supports positive potential double layers when the ratio of the average thermal velocity of positrons to that of electrons is less than a critical value. However, there exists a parameter regime for which the positive potential double layer is unable to restrict the occurrence of positive potential solitary waves and in this region of the parameter space, there exist positive potential solitary waves after the formation of a positive potential double layer. Consequently, positive potential supersolitons have been observed. The nonthermality of electrons plays an important role in the formation of positive potential double layers as well as positive potential supersolitons. The formation of positive potential supersoliton is analysed with the help of phase portraits of the dynamical system corresponding to the ion acoustic solitary structures of the present plasma system.
\end{abstract}



\section{Introduction}

Electron - positron plasmas along with significant amount of ions are found in early universe \citep{zeldovich1971,misner1973a,rees1983,weinberg72}, in active galactic nuclei \citep{miller1987,zurek1985}, around the pulsars \citep{shukla04}, in the remnants of supernova explosions \citep{alfven1981}, near the surface of the neutron stars \citep{zeldovich1971,shukla04}, and in the inner region of accretion discs in the vicinity of black holes \citep{lee2005,dubinov09}. The collission of primary cosmic rays with the neutral atoms of the uppermost atmosphere results in an amount of positrons \citep{gusev2001,ackermann2012}. Being trapped by the magnetic field of the Earth, these positrons are accumulated in the innermost magnetosphere of the Earth \citep{gusev2000}. Again, along with such positrons significant amount of ions are observed in the magnetosphere and in the lower ionosphere of the Earth \citep{alfven1981}. Therefore, electron-positron-ion (e-p-i) plasma may be present in the magnetosphere and in the lower ionosphere. The same type of nuclear interactions are naturally valid for the production of energetic antiparticles in the magnetosphere of other magnetized planets of our solar system \citep{gusev2001}. Therefore, the e-p-i plasma may also be found in the magnetosphere of Jupiter, Saturn and other magnetized planets. Such e-p-i plasma can be produced in laboratory \citep{liang1998,berezhiani1992pair,dubinov09}, like in tokamak plasmas \citep{helander2003,dubinov09} and in electron-positron-beam plasma experiment \citep{greaves1995,dubinov09}.

\cite{popel95} considered the nonlinear propagation of arbitrary amplitude ion acoustic (IA) waves in a plasma consisting of cold ions, isothermal electrons and positrons and they found the existence of compressive solitons only. \cite{nejoh96} investigated the large amplitude solitary waves in a plasma consisting of isothermal electrons, positrons, and positive ions and reported that the region of the existence of ion-acoustic solitons spreads as the the positron density and positron temperature increases. Using the  Bernoulli pseudo-potential method, \cite{dubinov09} determined the parameter regime for the existence of both periodic and solitary waves in an e-p-i plasma. \cite{shah2011} investigated the IA shock waves in a plasma consisting of relativistic adiabatic ions, kappa distributed electrons and positrons. \cite{baluku11} investigated the arbitrary amplitude IA solitary structures in a plasma consisting of cold ions, isothermal positrons and nonthermal electrons due to \cite{cairns95}. 
 
Motivated by the observations of solitary structures with density depletion made by Freja satellite \citep{dovner1994}, \cite{cairns95} have shown that the presence of nonthermal electrons change the properties of ion-sound solitary waves and solitons with both positive and negative density perturbations can exist. Following \cite{cairns95}, the number density of nonthermal energetic electrons can be written as
\begin{eqnarray}
\frac{n_{e}}{n_{e0}} = \bigg(1-\beta_{e}\frac{e\phi}{K_{B}T_{e}} +\beta_{e}\frac{e^{2}\phi^{2}}{K_{B}^{2}T_{e}^{2}}\bigg)
\exp\bigg[\frac{e\phi}{K_{B}T_{e}}\bigg],\nonumber \label{5}
\end{eqnarray}
where $\beta_{e} = 4 \alpha_{e}/(1 + 3 \alpha_{e})$ with $\alpha_{e} \geq 0$, $n_{e0}$ is the equilibrium number density of electrons, $T_{e}$ is the average temperature of electrons, $\phi$ is the electrostatic potential with $K_{B}$ is the Boltzmann constant. Here $\alpha_{e}$ and consequently, $\beta_{e}$ is the nonthermal parameter that determines the proportion of the fast energetic particles. Using the inequality $\alpha_{e} \geq 0$, it is easy to prove that $0 \leq \beta_{e} < 4/3$. However, we cannot take the whole region of $\beta_{e}$ ($0\leq \beta_{e}< 4/3$). Plotting the nonthermal velocity distribution against its velocity ($v$) in phase space, it can be easily shown that for increasing $\beta_{e}$, distribution function develops wings, which become stronger as $\beta_{e}$ increases. At the same time the center density in phase space drops, consequently, we should not take values of $\beta_{e} > 4/7$, since that stage might stretch the credibility of the Cairns model too far \citep{verheest08}. So, the effective range of $\beta_{e}$ is $0\leq \beta_{e} \leq  0.571429 \approx 0.6$. Such nonthermal electrons are observed in a number of astrophysical environments viz., in and around the Earth's Bow shock and Foreshock \citep{ABS68,FABGGMTPH83}, in the lower part of the magnetosphere of the Earth \citep{bostrom1992observations}. From the observation of Voyager 2, non-maxwellian distribution of electrons is expected in the magnetosphere of Saturn \citep{verheest00} and in the atmosphere of Uranus \citep{verheest00}. Beside this, \cite{LZPBHPDBLK89} reported the loss of energetic ions from the upper martian atmosphere and energetic protons are observed in the vicinity of the Moon \citep{FMSMH03}. Therefore, it is important to investigate the nonlinear wave structures in a plasma in which lighter species is nonthermally distributed. Specifically, in some cosmic sites, according to the prescription of \cite{alfven1981}, the velocity distribution function of charged particles are not only non-maxwellian but also highly anisotropic with an excess of high energy particles. In the present paper, Cairns model for the non-maxwellian distribution of electrons is taken into account whereas the velocity distribution of positrons is isothermal. It is also important to consider the nonthermal distribution of positrons but for simplicity and to see the effect of nonthermal electrons only, we take Maxwellian distribution of isothermal positrons. In a next paper, we shall consider the effect of both nonthermal electrons and nonthermal positrons on the ion acoustic solitary structures.

Again, the positrons present in a plasma system have a tendency to annihilate with electrons rapidly resulting the disappearance of positrons from the system. However, because  of  the  long lifetime of positrons, most of the astrophysical and laboratory plasmas can be considered as an admixture of electrons, positrons, and ions \citep{sabry09POP}. \cite{surko90} reported that even at an electron density of $1\times10^{12}$  $cm^{-3}$ and a temperature as low as 1 eV, the positron annihilation time is greater than 1 sec. In a laboratory environment (plasma temperature $300K$, characteristic dimension 10cm, volume $10^{3}$ $cm^{3}$), they became able to accumulate and store $\approx 1\times10^{9}$ $e^{+}$ for more than $10^{3}$ sec with Deby length of 2 mm and suggested that such a plasma would then be sufficiently large that a wide variety of plasma physics experiments could be considered.

In ion acoustic time scale, the characteristic wave frequency is very high, and consequently, the inevrse of the characteristic frequency is very small. Therefore, from the report of \cite{surko90}, we can expect that the positron annihilation time is larger than the inverse of the characteristic frequency of the ion-acoustic wave. Under this assumption (condition) the annihilation of positrons with electrons is negligible and the effect of positron annihilation can be neglected \citep{mishra07PRE,sabry09POP,sabry09PRE,jain13ASS}.

\cite{baluku11} reported the existence of negative potential double layer and also found the coexistence of solitary waves of both polarities. But the existence of arbitrary amplitude positive potential double layer in such plasma system has not been reported in literature. In the present work we consider the same plasma system of \cite{baluku11}, but with the only exception that the equation of pressure for ion fluid is taken into account to include the effect of ion temperature. For the present problem, we observe the following facts: (1) whenever the average temperature of positrons ($T_{p}$) is equal to the average temperature of electrons ($T_{e}$) then there is no qualitative change regarding the existence of IA solitary structurs as reported by \cite{baluku11} for any physically admissible value of ion temperature. (2) If we take $T_{p}<T_{e}$, then we get qualitatively new results regarding the nature of existence of different solitary structures, specifically, for $\sigma_{pe}< \sigma_{pe}^{(c)}$ ($\sigma_{pe}=T_{p}/T_{e}$ and $\sigma_{pe}^{(c)}$ is a cut-off value of $\sigma_{pe}$), we have found the exsitence of positive potential double layer (PPDL) and positive potential supersoliton (PPSS) for small values of the positron concentration. Investigations of the positive potential double layers (PPDLs) and the positive potential supersolitons in this e-p-i plasma system have not been considered in the literature. \cite{dubinov12a,dubinov12b} have recently introduced a new type of solitary structure which they termed as `supersoliton'. After this work several authors \citep{dubinov12c,das12,verheest13a,verheest13b,verheest13c,hellberg13,maharaj13,verheest14,verheest14a,lakhina14,singh15,verheest2015,paul2016} reported the existence of supersolitons in different plasma systems. Although, a number of investigations \citep{eslami11POP,jain13ASS,jain13jpp,ghosh14} on IA solitary structures in different e-p-i plasma has been carried out but the existence of IA supersolitons in those plasma systems has not been reported. In the present paper, we have also investigated the formation of positive potential supersoliton with the help of phase portraits of the dynamical system corresponding to the ion acoustic solitary structures. 

The present paper is organized as follows: the basic equations are given in \S \ref{sec:basic_eqn}. The derivation and the mechanical analogy of the energy integral have been presented in \S \ref{sec:energy_int}. In \S \ref{sec:solution_spaces}, IA solitary structures have been discussed with the help of the qualitatively distinct compositional parameter spaces. Phase portraits of the dynamical system corresponding to the ion acoustic solitary structures have been analyzed in \S \ref{sec:phase_portrait} giving special emphasis on positive potential supersolitons.  Finally, conclusions are given in \S \ref{sec:Summary_Discussions}.

\section{\label{sec:basic_eqn}Basic Equations}
The following are the governing equations describing the nonlinear behaviour of IA waves  propagating  along  $x$-axis  in  collisionless  unmagnetized   plasma  consisting of adiabatic warm ions, nonthermal electrons and isothermal positrons:
%
\begin{eqnarray}\label{continuity}
\frac{\partial n_{i}}{\partial t}+\frac{\partial}{\partial x}(n_{i}u_{i})=0,
\end{eqnarray}
\begin{eqnarray}\label{momentum}
M_{s}^{2}\bigg(\frac{\partial u_{i}}{\partial t}+u_{i}\frac{\partial u_{i}}{\partial x}\bigg)+\frac{(1-p)\sigma_{ie}}{n_{i}}\frac{\partial p_{i}}{\partial x}+\frac{\partial \phi}{\partial x}=0,
\end{eqnarray}
\begin{eqnarray}\label{pressure}
\frac{\partial p_{i}}{\partial t}+u_{i}\frac{\partial p_{i}}{\partial x}+\gamma p_{i} \frac{\partial u_{i}}{\partial x}=0,
\end{eqnarray}
\begin{eqnarray}\label{poisson}
\frac{\partial^{2} \phi}{\partial x^{2}}=-\frac{M_{s}^{2}-\gamma \sigma_{ie}}{1-p}(n_{i}-n_{e}+n_{p}).
\end{eqnarray}
Here $ n_{i}$, $n_{e}$, $n_{p}$, $u_{i}$, $p_{i}$, $\phi$, $x $ and $ t $ are, respectively, ion number density, electron number density, positron number density, ion fluid velocity, ion fluid pressure, electrostatic potential, spatial variable and time, and these field variables have been normalized by $n_{e0}$ ($=n_{i0}+n_{p0}$), $n_{e0}$, $n_{e0}$, $C_{S}$ (linearized velocity of the IA wave in the present plasma system for long-wave length plane wave perturbation), $n_{i0}K_{B}T_{i}$, $\Phi=\frac{K_{B}T_{e}}{e}$, $ \lambda_{D} $ (Debye length of the present plasma system) and $T=\frac{\lambda_{D}}{C_{S}}$ respectively with $ \gamma(=3) $ is the adiabatic index, $n_{e0}$, $n_{i0}$ and $n_{p0}$ are respectively the unperturbed electron, ion and positron number densities, $K_{B}$ is the Boltzmann constant. The expressions of $M_{s}$ and of the parameters $p$, $\sigma_{ie}$, $\sigma_{pe}$ are given by
\begin{eqnarray}\label{Ms}
M_{s}=\sqrt{\gamma\sigma_{ie}+\frac{(1-p)\sigma_{pe}}{p+(1-\beta_{e}) \sigma_{pe}}},
\end{eqnarray}
\begin{eqnarray}\label{p}
p=\frac{n_{p0}}{n_{e0}},~\sigma_{ie}=\frac{T_{i}}{T_{e}},~\sigma_{pe}=\frac{T_{p}}{T_{e}},
\end{eqnarray}
where $T_{i}$ is the average temperature of ions.

The equations (\ref{continuity}) - (\ref{poisson}) are supplemented by the following normalized number densities of nonthermal electrons and isothermal positrons
\begin{eqnarray}\label{ne}
n_{e} = (1-\beta_{e}\phi+\beta_{e}\phi^{2})e^{\phi},~n_{p} = p e^{-\phi / \sigma_{pe}}, 
\end{eqnarray}
together with the charge neutrality condition : $ n_{i0}+n_{p0}=n_{e0} $.

\section{\label{sec:energy_int} Energy Integral}
To  study  the  arbitrary  amplitude  time  independent IA solitary structures, we suppose that all the dependent variables depend only on a single variable $ \xi=x-Mt $, where $ M $ is independent of $ x $ and $ t $. Therefore, lifting the equations (\ref{continuity})-(\ref{poisson}) in the wave frame moving with a constant velocity $M$ normalized by the linearized IA speed ($C_{S}$), using the boundary conditions: $ \big(n_{i},p_{i},u_{i},\phi,\frac{d\phi}{d\xi}\big)\rightarrow \big(1-p,1,0,0,0\big)\mbox{    as    }  |\xi|\rightarrow \infty,
$ and following the method of \cite{das12,das09}, we get the following energy integral:
\begin{eqnarray}\label{energy_integral}
\frac{1}{2}\bigg(\frac{d\phi}{d\xi}\bigg)^{2}+V(\phi)=0,
\end{eqnarray}
where
\begin{eqnarray}\label{V_phi_1}
V(\phi) = (M_{s}^{2}-3\sigma_{ie}) \Big[V_{i}+\frac{p }{1-p} \sigma_{pe} V_{p}-\frac{1}{1-p}V_{e}\Big],
\end{eqnarray}
\begin{eqnarray}\label{V_i_1}
V_{i} &=& M^{2}M_{s}^{2}+\sigma_{ie}\nonumber \\ &&-N_{i}(M^{2}M_{s}^{2}+3\sigma_{ie}-2\phi-2\sigma_{ie}N_{i}^{2}),
\end{eqnarray}
\begin{eqnarray}\label{N_i_1}
N_{i}=\frac{n_{i}}{1-p}=\frac{MM_{s}\sqrt{2}}{(\sqrt{\Phi_{M}-\phi}+\sqrt{\Psi_{M}-\phi})}
\end{eqnarray}
\begin{eqnarray}\label{Phi_M_1}
\Phi_{M} &=& \frac{1}{2}\bigg(MM_{s}+\sqrt{3\sigma_{ie}}\bigg)^{2},
\end{eqnarray}
\begin{eqnarray}\label{Psi_M_1}
\Psi_{M} &=& \frac{1}{2}\bigg(MM_{s}-\sqrt{3\sigma_{ie}}\bigg)^{2},
\end{eqnarray}
\begin{eqnarray}\label{V_e_1}
V_{e} &=& (1+3\beta_{e}-3\beta_{e}\phi+\beta_{e}\phi^{2})e^{\phi}-(1+3\beta_{e}),
\end{eqnarray}
\begin{eqnarray}\label{V_p_1}
V_{p} &=& 1-e^{-\phi/\sigma_{pe}}.
\end{eqnarray}

Using the mechanical analogy, \cite{sagdeev66} established that for the existence of a positive (negative) potential solitary wave [PPSW] ([NPSW]) solution of (\ref{energy_integral}), the following three conditions must be simultaneously satisfied:
(i) $\phi=0$ is the position of unstable equilibrium of a particle of unit mass associated with the energy integral (\ref{energy_integral}), i.e., $V(0)=V'(0)=0$ and $V''(0)<0$.
(ii) $V(\phi_{m}) = 0$, $V'(\phi_{m}) > 0$ ($V'(\phi_{m}) < 0$) for some $\phi_{m} > 0$ ($\phi_{m} < 0$). This condition is responsible for the oscillation of the particle within the interval $\min\{0,\phi_{m}\}<\phi<\max\{0,\phi_{m}\}$.
(iii) $V(\phi) < 0$ for all $0 <\phi < \phi_{m}$ ($\phi_{m} < \phi < 0$). This condition is necessary to define the energy integral (\ref{energy_integral}) within the interval $\min\{0,\phi_{m}\}<\phi<\max\{0,\phi_{m}\}$. For the existence of a positive (negative) potential double layer [PPDL] ([NPDL]) solution of (\ref{energy_integral}), the second condition is replaced by $V(\phi_{m}) = 0$, $V'(\phi_{m}) = 0$, $V''(\phi_{m}) < 0$ for some $\phi_{m} > 0$ ($\phi_{m} < 0)$, which states that the particle cannot be reflected back from the point $\phi=\phi_{m}$ to the point $\phi = 0$.

The necessary conditions for  the  existence  of  solitary structures  of  the  energy  integral  (\ref{energy_integral}) give $M>M_{c}=1$, i.e., the solitary structures start to exist for $M > M_{c}=1$. However, we have seen in the literature \citep{das12mc} that if there is a coexistence of PPSWs and NPSWs in some region of the parameter space, then one of the two polarities has a finite amplitude soliton at the minimum acoustic speed, which is the acoustic speed corresponding to the mach number $M=M_{c} = 1$. Here, we have not discussed the case for the existence of solitary structures when $V''(0)=0$. Following \cite{das12mc} one can discuss this case for the present problem.

From the expression of $ N_{i} $ as given by (\ref{N_i_1}), we see that $ N_{i} $ is  well - defined  if and only if $ \phi \leq \Psi_{M} $.  Using the condition $ \phi \leq \Psi_{M} $ and following \cite{das09,das12}, it is simple to check that for the existence of all PPSWs, the  mach  number $ M $ is  restricted  by  $ M_{c} < M \leq M_{max} $, where $ M_{max} $ is the largest positive root of equation $ V(\Psi_{M}) = 0 $ subject to the condition $ V(\Psi_{M}) \geq 0 $ for all $ M \leq M_{max} $. So, $M$ assumes its upper limit $ M_{max} $ for the existence of all PPSWs when $\phi$ tends to $\Psi_{M}$, i.e., when ion number density goes to maximum compression.


Following the subsections \S 5.3 and \S 5.4 of section \S 5 of \cite{das12} one can easily develop an algorithm to find the mach number $M_{NPDL}$ ($M_{PPDL}$) corresponding to a negative (positive) potential double layer solution of the energy integral (\ref{energy_integral}).

Now, double layer solution restricts the occurrence of at least one sequence of solitary waves of same polarity. We have seen that $M_{max}$ can restrict the existence of all PPSWs. Now if both $M_{max}$ and $M_{PPDL}$ exist, then we must have $M_{c}<M_{PPDL} < M_{max}$ and we can split the entire range of $M$ into two disjoint subintervals, viz., $M_{c}<M<M_{PPDL}$ and $M_{PPDL}<M \leq M_{max}$. For $M_{c}<M<M_{PPDL}$, we get a sequence of PPSWs converging to the PPDL solution at $M=M_{PPDL}$ whereas for $M_{PPDL}<M \leq M_{max}$, we get PPSWs after the formation of PPDL, and consequently, the existence of positive potential supersoliton (according to \cite{dubinov12a}) is confirmed. However, Verheest and co-workers \citep{verheest13a,verheest13b,verheest13c,verheest14,verheest14a} plotted the existence diagrams for a number of different plasma configurations, and showed that the lower limit of the mach number for the existence of supersolitons could be: (i) greater than the speed of double layer,  (ii) exactly at the acoustic speed or  (iii) the structure speed at which a subsidiary maximum first develops within the pseudopotential well. In the present problem, we consider the supersoliton structures that occur beyond double layers with the help of qualitatively different existence domains. 

\section{\label{sec:solution_spaces} Different existence domains}
Figure \ref{sol_spc_wrt_beta_e_p=0} - Figure \ref{sol_spc_wrt_beta_e_sigma_pe=0_pt_2} are different existence domains or compositional parameter spaces showing the nature of existence of different solitary structures. We have made the following general descriptions to explain those figures. Solitary structures start to exist just above the lower curve $ M = M_{c} =1 $. At each point on the curve $M=M_{PPDL}$ ($M=M_{NPDL}$), one can get a PPDL (NPDL) whereas at each point on the curve $M=M_{max}$, one can get a PPSW. In absence of $M_{PPDL}$ ($ M_{max}$), $ M_{max}$ ($ M_{PPDL}$) is the upper bound of $ M $ for the existence of PPSWs.
If both $M_{PPDL}$ and $M_{max}$ exist finitely, then $\max\{M_{max},M_{PPDL}\}$ is the upper bound of $ M $ for the existence of PPSWs.
If we pick a $\beta_{e}$ and go vertically upwards, then all intermediate values of $ M $ bounded by the curves $ M=M_{c} $ and $ M=M_{max} $ or $M_{PPDL}$ or $\max\{M_{max},M_{PPDL}\}$ would give PPSWs.
Similarly, all intermediate values of $ M $ bounded by the curves $ M=M_{c} $ and $ M=M_{NPDL} $ would give NPSWs.
Now, if we can find a region where the existence region of PPSWs (NPSWs) is separated by the curve $M=M_{PPDL}$ ($M=M_{NPDL}$), then we can claim that the system supports positive (negative) potential supersolitons.
In particular, if both $M_{PPDL}$ and $M_{max}$ exist and $M_{PPDL}<M_{max}$ then all intermediate values of $ M $ bounded by the curves $ M=M_{PPDL} $ and $ M=M_{max} $ would give positive potential supersolitons. For $M_{PPDL}<M \leq M_{max}$, amplitude of the positive potential supersoliton increases with increasing values of $M$ and the maximum amplitude of the supersoliton can be obtained at $M=M_{max}$. We have used the following notations : C -- Region of coexistence of both NPSWs and PPSWs, N -- Region of existence of NPSWs, P -- Region of existence of PPSWs and S -- Region of existence of positive potential supersolitons.

Figure \ref{sol_spc_wrt_beta_e_p=0}, figure \ref{sol_spc_wrt_beta_e_p=0_pt_002}, figure \ref{sol_spc_wrt_beta_e_p=0_pt_005} and figure \ref{sol_spc_wrt_beta_e_p=0_pt_02} correspond to the existence domains for different values of positron concentration $p$ starting from $p=0$ with $\sigma_{ie}=0.01$ and $\sigma_{pe}=0.1$. Although the figure \ref{sol_spc_wrt_beta_e_p=0} is the existence domain for $p=0$ but qualitatively it represents the compositional parameter space for any $p$ lying within $0 \leq p \leq 0.0012$. Similarly, figure \ref{sol_spc_wrt_beta_e_p=0_pt_002} stands for any $p$ lying within $0.0012 < p \leq 0.0035$ whereas figure \ref{sol_spc_wrt_beta_e_p=0_pt_005} and figure \ref{sol_spc_wrt_beta_e_p=0_pt_02} represent the existence domains for any $p$ lying within $0.0035 < p \leq 0.0136$ and $p>0.0136$ respectively. It is important to note that any smaller value of $\sigma_{ie}$ including zero gives qualitatively same existence domain. 

According to the general description of the existence domain, figure \ref{sol_spc_wrt_beta_e_p=0}, figure \ref{sol_spc_wrt_beta_e_p=0_pt_002}, figure \ref{sol_spc_wrt_beta_e_p=0_pt_005} and figure \ref{sol_spc_wrt_beta_e_p=0_pt_02} are self explanatory. For example, consider figure \ref{sol_spc_wrt_beta_e_p=0_pt_002}. From this figure we have the following observations. (i) The system supports PPSW up to a cut-off value $\beta_{e}^{(c)}$ of $\beta_{e}$ for all $M$ lying within $M_{c}<M \leq M_{max}$ but $M \neq M_{PPDL}$. The system supports PPDL along the curve $M=M_{PPDL}$. (ii) There exist two cut-off values $\beta_{e}^{(a)}$ and $\beta_{e}^{(b)}$ such that for $\beta_{e}^{(a)}<\beta_{e}<\beta_{e}^{(b)}$, we have $M_{PPDL}<M_{max}$, i.e., in this interval of $\beta_{e}$, there exist PPSWs after the formation of double layers if the mach number $M$ is restricted by $M_{PPDL}<M \leq M_{max}$ and consequently, in this region of existence domain, the existence of positive potential supersolitons is confirmed. (iii) The system supports NPDL along the curve $M=M_{NPDL}$. (iv) NPSWs start to exist if $\beta_{e}$ exceeds a cut-off value $\beta_{e}^{(d)}$ and the mach number $M$ is restricted by $M_{c}<M<M_{NPDL}$. (v) The system does not support any negative potential supersoliton.

Figure \ref{sol_spc_wrt_beta_e_p=0_pt_002} and figure \ref{sol_spc_wrt_beta_e_p=0_pt_005} confirm the existence of positive potential supersoliton in a certain interval of $\beta_{e}$ for $p$ lying in the interval $p^{(a)}<p \leq p^{(c)}$, where $p^{(a)}$ and $p^{(c)}$ are two cut-off values of $p$. The existence region of positive potential supersoliton increases with the increase in the values of $p$ up to a certain cut-off value of $p$ and as $p$ exceeds this value, the existence region of positive potential supersoliton decreases for increasing values of positron concentration $p$ and finally, for
$p>p^{(c)}$ the system does not support any positive potential supersoliton.

It is easy to check that for fixed values of $p$ and $\sigma_{pe}$ the existence region of positive potential supersoliton decreases with the increase in the values of $\sigma_{ie}$ and there exist a cut-off value of $\sigma_{ie}$ for which the system does not support any positive potential supersoliton.

Again, it is simple to verify that for fixed values of $p$ and $\sigma_{ie}$, the existence region of positive potential supersoliton increases with the increase in the values of $\sigma_{pe}$ until $\sigma_{pe}$ takes the cut-off value $\sigma_{pe}^{(b)}$, and for $\sigma_{pe}>\sigma_{pe}^{(b)}$ the existence region of positive potential supersoliton decreases very rapidly and finally, there exists a critical value $\sigma_{pe}^{(c)}$ of $\sigma_{pe}$ such that the system does not support any positive potential supersoliton for $\sigma_{pe}>\sigma_{pe}^{(c)}$. In fact, for $\sigma_{pe}>\sigma_{pe}^{(c)}$, the existence domain is qualitatively the same as given in figure \ref{sol_spc_wrt_beta_e_p=0}. Figure \ref{sol_spc_wrt_beta_e_sigma_pe=0_pt_2}(a) shows the existence domain for $\sigma_{pe}=0.2(>\sigma_{pe}^{(c)})$ for fixed $p$ and $\sigma_{ie}$. For $\sigma_{ie}=0.01$ and $p=0.005$ the value of $\sigma_{pe}^{(b)}$ is 0.19 and the value of $\sigma_{pe}^{(c)}$ is very close to the value of $\sigma_{pe}^{(b)}$.

Another interesting observation is that if we fix the values of $\sigma_{ie}$ and $\sigma_{pe}$ as described in figure \ref{sol_spc_wrt_beta_e_sigma_pe=0_pt_2}(a) but increase the concentration of positron to $p=0.015$ and draw the existence domain with respect to $\beta_{e}$, then this existence domain shows that the system supports positive potential supersoliton but does not support any NPSW. Figure \ref{sol_spc_wrt_beta_e_sigma_pe=0_pt_2}(b) describes the existence domain for $p=0.015$, $\sigma_{ie}=0.01$ and $\sigma_{pe}=0.1$.

\section{\label{sec:phase_portrait} Phase portraits of the dynamical system corresponding to the IA solitary structures}

Differentiating the energy integral (\ref{energy_integral}) with respect to $\phi$, we get
\begin{eqnarray}\label{energy_integral_differentiation}
\frac{d^{2}\phi}{d\xi^{2}}+V'(\phi)=0.
\end{eqnarray}
This equation is equivalent to the following system of differential equations
\begin{eqnarray}\label{phase_portraits}
\frac{d\phi_{1}}{d\xi}=\phi_{2},~\frac{d\phi_{2}}{d\xi}=-V'(\phi_{1}),
\end{eqnarray}
where $\phi_{1}=\phi$. In the previous section, we have observed the supersoliton structures that occur beyond double layers. Now, we explain their different unusual shapes with the help of phase portrait of the system of coupled equations (\ref{phase_portraits}) in the $\phi_{1}-\phi_{2}$ plane.

To describe the existence and the shape of positive potential supersolitons and the coexistence of solitons of both polarities, we consider Figure \ref{anup_PPSW_phase_portrait_phi_Vphi_new} - Figure \ref{supersoliton_fixed_points}, where we have used the existence domain as shown in Figure \ref{sol_spc_wrt_beta_e_p=0_pt_002} to determine the mach numbers for the formation of positive potential supersolitons. The existence domain for $p=0.000001$, $\beta_{e}=0.47$, $\sigma_{ie}=0.01$ and $\sigma_{pe}=0.1$ has been considered to determine the mach numbers for coexistence of solitons of both polarities.

In Figure \ref{anup_PPSW_phase_portrait_phi_Vphi_new} - Figure \ref{another_pp_supersoliton}, $V(\phi)$ is plotted against $\phi$ in the upper panel (or marked as (a)) of each figure. The lower panel (or marked as (b)) of each figure shows the phase portrait of the system (\ref{phase_portraits}). In the above mentioned figures, we have used the values of the parameters as indicated in the figures with $\sigma_{pe}=0.1$ and $\sigma_{ie}=0.01$. The curve $V(\phi)$ and the  phase portrait have been drawn on the same horizontal axis $\phi(=\phi_{1})$. Small solid circle corresponds to a saddle point whereas the small solid star indicates an equilibrium point other than saddle point of the system (\ref{phase_portraits}). It is simple to check that each maximum (minimum) point of $V(\phi)$ corresponds to a saddle point (an equilibrium point other than a saddle point) of the system (\ref{phase_portraits}).

From these figures, we see that there is a one-one correspondence between the separatrix of the phase portrait as shown with a heavy bold line in the lower panel with the curve $V(\phi)$ against $\phi$ of the upper panel. Again, it is important to note that the origin $(0,0)$ is always a saddle point of the system (\ref{phase_portraits}) and the separatrix corresponding to a solitary structure appears to start and end at the saddle point (0,0). The separatrix corresponding to a solitary structure is shown with a heavy bold line whereas other separatrices (if exist) are shown by bold lines. The closed curve about an equilibrium point (other than a saddle point) contained in at least one separatrix indicates the possibility of the periodic wave solution about that fixed point. For example, the closed curves of Figure \ref{anup_PPSW_phase_portrait_phi_Vphi_new}(b) about the fixed point (0.0172,0) lying within the separatrix indicate the possibility of the periodic wave solutions about the fixed point (0.0172,0).

Figure \ref{anup_PPSW_phase_portrait_phi_Vphi_new}(a) shows the existence of a PPSW before the formation of PPDL whereas the Figure \ref{anup_PPSW_phase_portrait_phi_Vphi_new}(b) shows that the corresponding phase portrait contains only one saddle at the origin and a non-zero eqilibrium point. Consequently, there exists only one separatrix that appears to start and end at the origin enclosing a non-saddle fixed point. More precisely, the trajectory corresponding to the separatrix approaches the origin as $\xi \rightarrow \pm \infty$. It is also important to note that a separatrix corresponding to a solitary structure does not correspond to a periodic solution because for this case, the trajectory takes forever trying to reach a saddle point. In fact, this is the reason that a pseudo-particle associated with the energy integral (\ref{energy_integral}) takes an infinite long time to move away from its unstable position of equilibrium and then it continues its motion until $\phi$ takes the value $\phi_{m} (>0)$, where $V(\phi_{m})=0$ and $V'(\phi_{m})>0$ and again it takes an infinite long time to come back its unstable position of equilibrium \citep{verheest00}. Similarly, Figure \ref{anup_Supersoliton_phase_portrait_phi_Vphi} confirms the existence of a PPSW after the formation of PPDL.

From the phase portraits as given in Figure \ref{anup_PPSW_phase_portrait_phi_Vphi_new}(b) and Figure \ref{anup_Supersoliton_phase_portrait_phi_Vphi}(b), we see that there is no qualitative difference between these two phase portraits. Again, according to \cite{dubinov12a}, the separatrix corresponding to a supersoliton envelopes one or several inner separatrices and several equilibrium points. So, according to \cite{dubinov12a}, Figure \ref{anup_Supersoliton_phase_portrait_phi_Vphi}(b) does not correspond to a supersoliton. But Figure \ref{profile_phi_vs_xi} shows that there is a finite jump between the amplitudes of solitons before and after the formation of double layer. To expain this fact, we first of all consider the phase portait corresponding to a double layer solution as given in Figure \ref{anup_PPDL_phase_portrait_phi_Vphi}(b). Figure \ref{anup_PPDL_phase_portrait_phi_Vphi}(b) shows that the separatrix corresponding to the double layer solution appears to pass through two saddle points and it encloses another two equilibrium points. If both the saddle points exist after a small increament of $M$ from $M=M_{PPDL}$ then the separatrix appears to pass through the origin encloses an inner sparatrix through a non-zero saddle and at least two equilibrium points as shown in the lower panel of Figure \ref{another_pp_supersoliton}. Therefore, according to \cite{dubinov12a}, we see that for the same set of values of the parameters, $M=M_{PPDL}+0.00005$ defines a supersoliton whereas $M=M_{PPDL}+0.0001$ does not define a supersoliton. But in both the cases we have a finite jump between the amplitudes of solitons after and before the formation of double layer. To make a clear difference between the solitons given in Figure \ref{anup_Supersoliton_phase_portrait_phi_Vphi}(b) and the lower panel of Figure \ref{another_pp_supersoliton} for $M=M_{PPDL}+0.0001$ and $M=M_{PPDL}+0.00005$ respectively, we consider Figure \ref{supersoliton_fixed_points}. In this figure, we draw the saddle and other eqilibrium points of the system (\ref{phase_portraits}) on the $\phi(=\phi_{1})$-axis for increasing values of $M$ starting from $M=M_{PPDL}+0.000001$. This figure shows that for increasing values of $M$ the distance between the non-zero saddle and the equilibrium point nearest to it decreases and ultimately both of them disappear from the system. Finally, the system contains only one saddle at the origin and a non-zero eqilibrium point. Consequently, only one separatrix enclosing the non-saddle fixed point is possible that appears to pass through the saddle at the origin. So, the existence of a soliton after the formation of a double layer confirms the existence of a sequence of supersolitons.

From the phase portrait of the coexistence of solitons of both polarities (lower panel of Figure \ref{anup_coexistence_phase_portrait_phi_Vphi}), we have the following observations: there are two separatrices and the separatrix corresponding to the coexistence of solitons of both polarities (shown with a heavy bold line) is contained in another separatrix (shown with a bold line). There exist infinitely many closed curves between these two separatrices and each of these closed curves corresponds to a super-nonlinear perodic wave as shown in Figure 5(c) and Figure 6(c) in the paper of \cite{dubinov12} for a dusty plasma system. However, further investigation of the super-nonlinear perodic wave solutions of the energy integral (\ref{energy_integral}) is beyond the scope of this paper. 

\section{\label{sec:Summary_Discussions} Summary \& Discussions }

We have made a thorough investigation on the nature of existence of different IA solitary structures in an unmagnetized colissionless plasma composed of adiabatic warm ions, nonthermal electrons and isothermal positrons with the help of qualitatively different existence domains. The coexistence of solitary waves of both polarities and the existence of NPDLs have been observed. These results are very much consistent with the existing literature. Here, we have found a parameter regime for which the system supports PPDLs. The system does not support any double layer solution for isothermal electrons. The nonthermality of electrons plays an important role in the formation of double layers of both polarities. The present system also supports positive potential supersolitons for nonthermal electrons only. But it does not support any negative potential supersoliton. Not only $\beta_{e}$ and $p$ but also the parameters $\sigma_{pe}$ and $\sigma_{ie}$ have significant role in the formation of positive potential supersolitons.

\begin{figure}
  \includegraphics{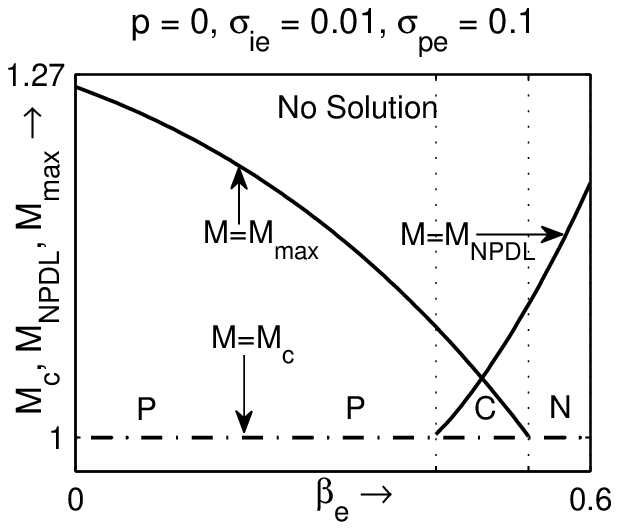}
  \caption{\label{sol_spc_wrt_beta_e_p=0} Compositional parameter space with respect to $\beta_{e}$ for $p=0$, $\sigma_{ie}=0.01$ and $\sigma_{pe}=0.1$}.
\end{figure}

\begin{figure}
  \includegraphics{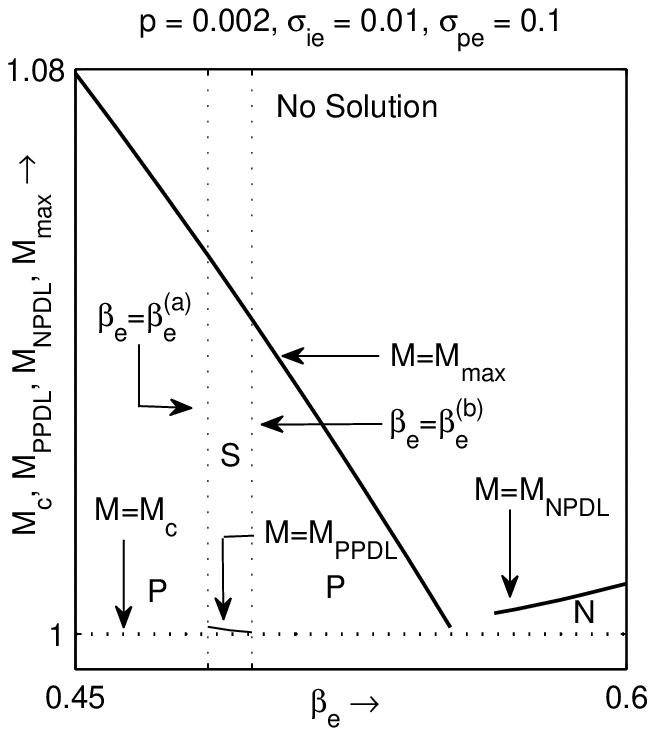}
  \caption{\label{sol_spc_wrt_beta_e_p=0_pt_002} Compositional parameter space with respect to $\beta_{e}$ for $p=0.002$, $\sigma_{ie}=0.01$ and $\sigma_{pe}=0.1$}.
\end{figure}

\begin{figure}
  \includegraphics{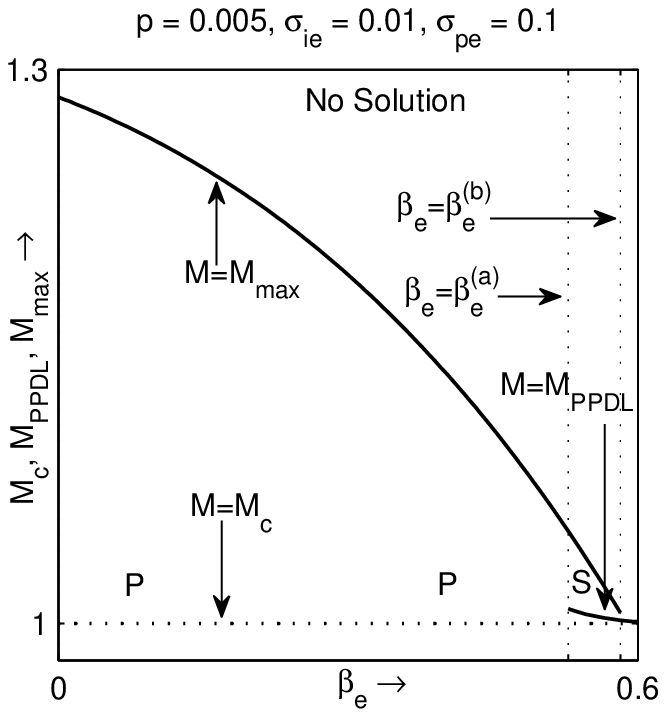}
  \caption{\label{sol_spc_wrt_beta_e_p=0_pt_005} Compositional parameter space with respect to $\beta_{e}$ for $p=0.005$, $\sigma_{ie}=0.01$ and $\sigma_{pe}=0.1$}.
\end{figure}

\begin{figure}
  \includegraphics{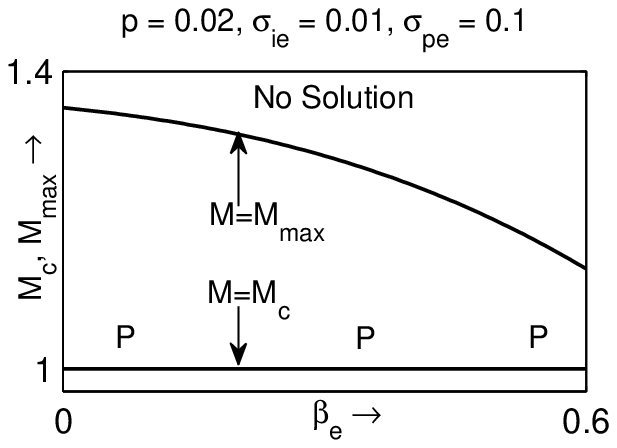}
  \caption{\label{sol_spc_wrt_beta_e_p=0_pt_02} Compositional parameter space with respect to $\beta_{e}$ for $p=0.02$, $\sigma_{ie}=0.01$ and $\sigma_{pe}=0.1$}.
\end{figure}

\begin{figure}
  \includegraphics{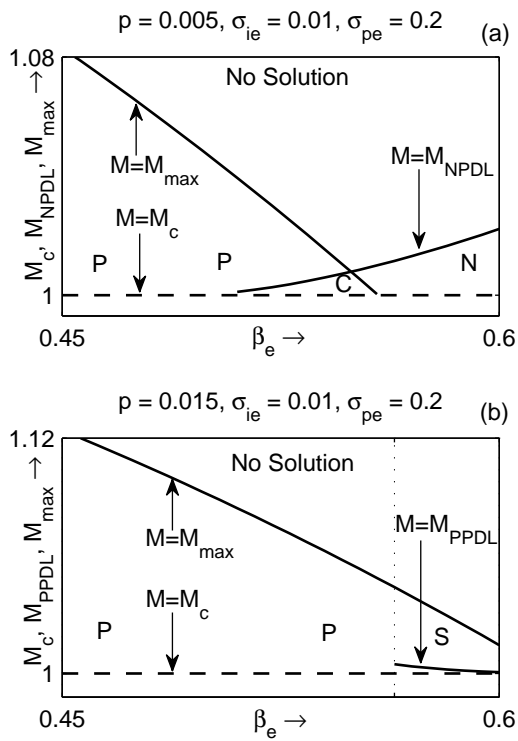}
  \caption{\label{sol_spc_wrt_beta_e_sigma_pe=0_pt_2} Compositional parameter space with respect to $\beta_{e}$ for $\sigma_{ie}=0.01$ and $\sigma_{pe}=0.2$ with (a) $p=0.005$,  (b) $p=0.015$.}
\end{figure}

\begin{figure}
  \includegraphics{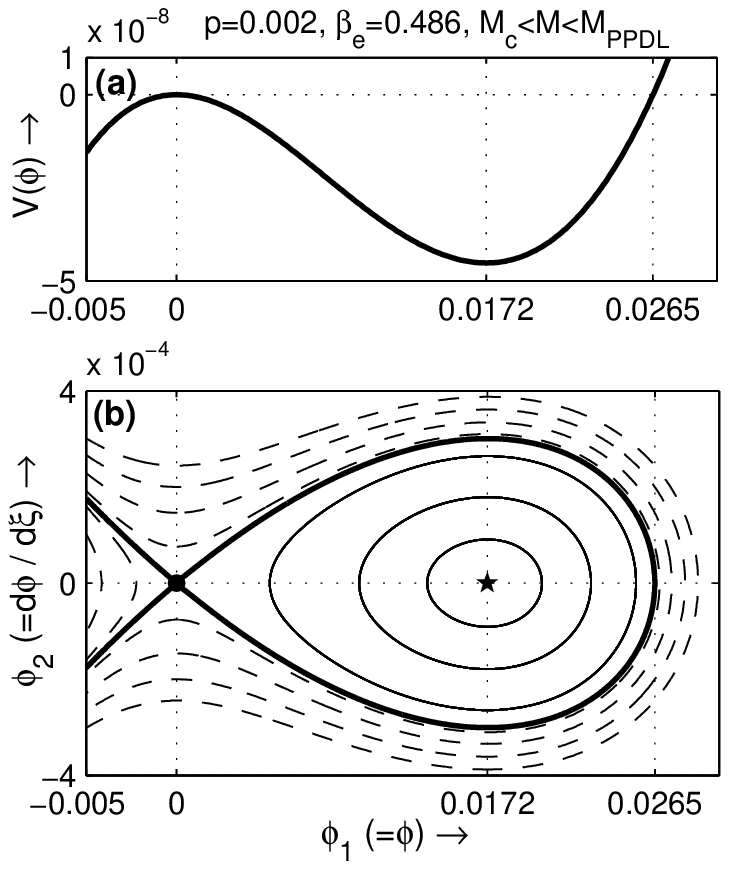}
  \caption{\label{anup_PPSW_phase_portrait_phi_Vphi_new} $V(\phi)$ (on top) and the phase portrait of the system (\ref{phase_portraits}) (on bottom) have been drawn on the same $\phi(=\phi_{1})$-axis for $M=M_{PPDL}-0.0001$.}
\end{figure}

\begin{figure}
  \includegraphics{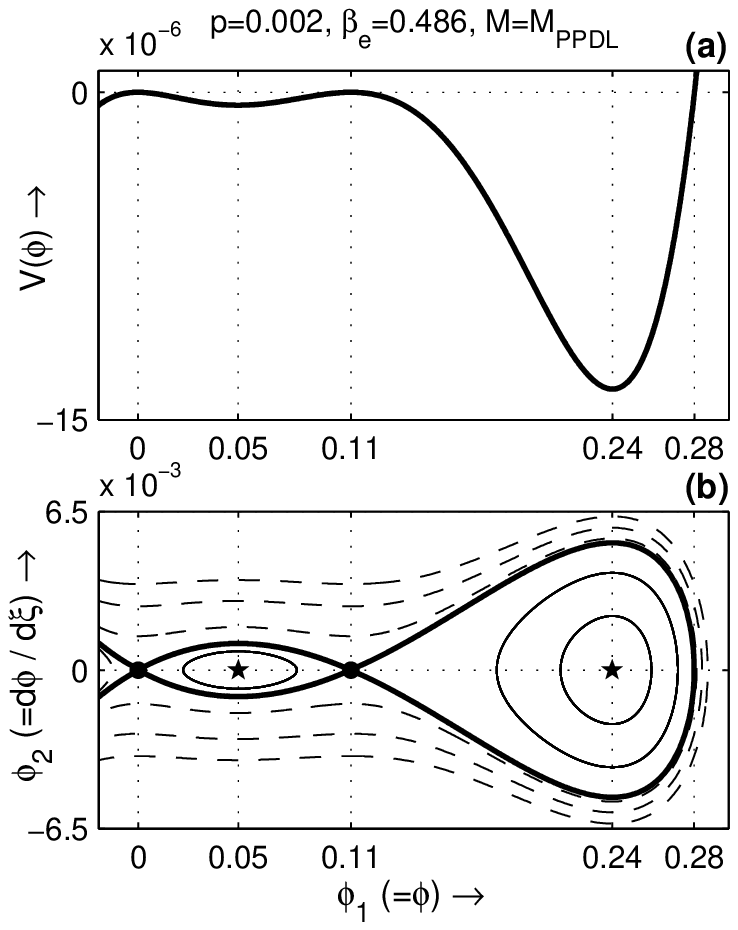}
  \caption{\label{anup_PPDL_phase_portrait_phi_Vphi} $V(\phi)$ (on top) and the phase portrait of the system (\ref{phase_portraits}) (on bottom) have been drawn on the same $\phi(=\phi_{1})$-axis for $M=M_{PPDL}$.}
\end{figure}

\begin{figure}
  \includegraphics{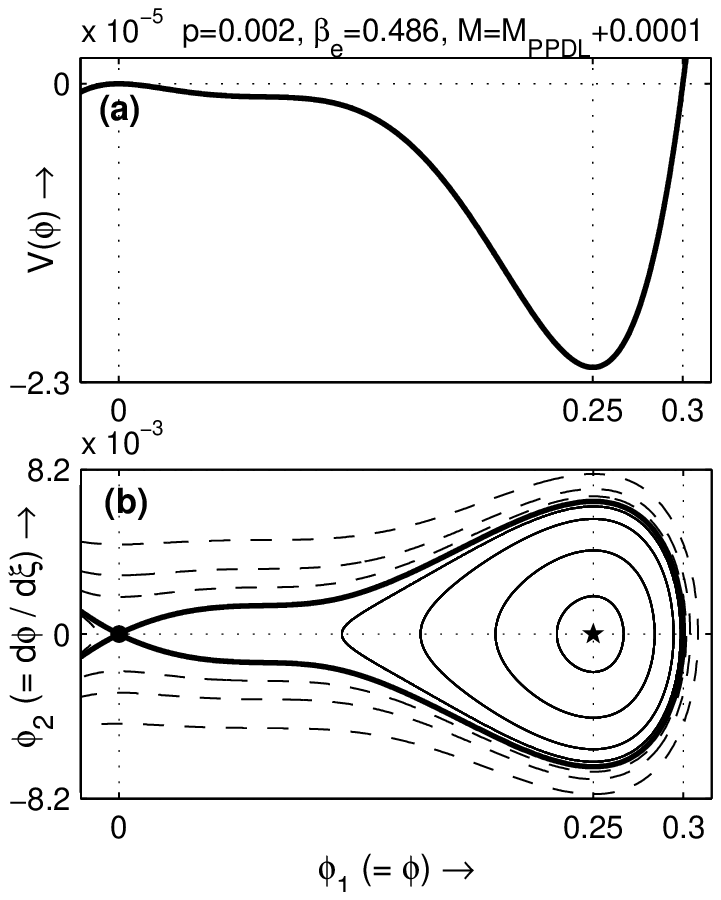}
  \caption{\label{anup_Supersoliton_phase_portrait_phi_Vphi} $V(\phi)$ (on top) and the phase portrait of the system (\ref{phase_portraits}) (on bottom) have been drawn on the same $\phi(=\phi_{1})$-axis for $M=M_{PPDL}+0.0001$.}
\end{figure}

\begin{figure}
  \includegraphics{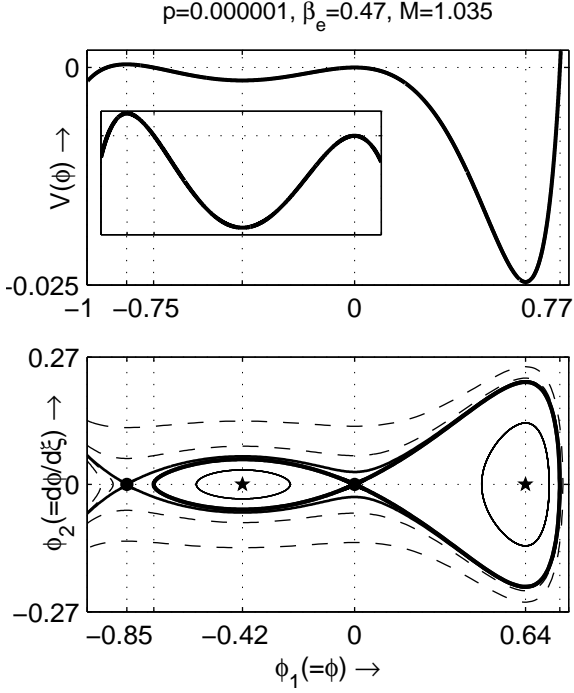}
  \caption{\label{anup_coexistence_phase_portrait_phi_Vphi} $V(\phi)$ (on top) and the phase portrait of the system (\ref{phase_portraits}) (on bottom) have been drawn on the same $\phi(=\phi_{1})$-axis corresponding to the coexistence of solitons of both polarities.}
\end{figure}

\begin{figure}
  \includegraphics{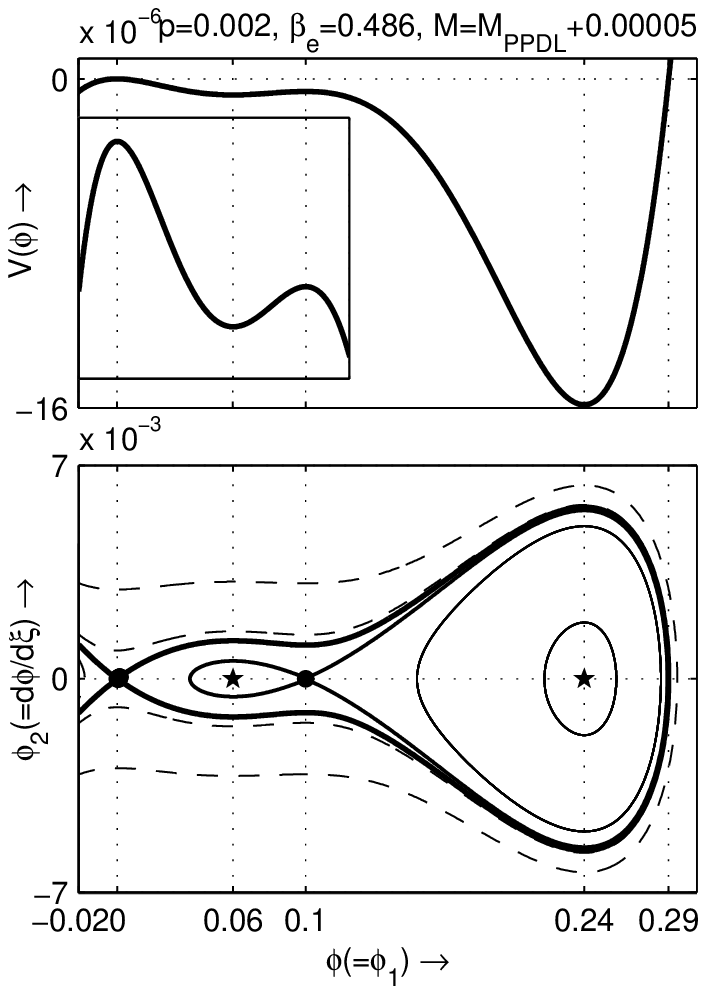}
  \caption{\label{another_pp_supersoliton} $V(\phi)$ (on top) and the phase portrait of the system (\ref{phase_portraits}) (on bottom) have been drawn on the same $\phi(=\phi_{1})$-axis for $M=M_{PPDL}+0.00005$.}
\end{figure}

\begin{figure}
  \includegraphics{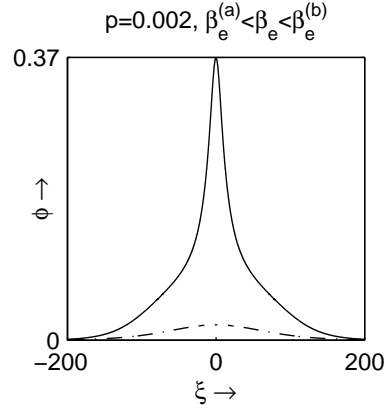}
  \caption{\label{profile_phi_vs_xi} $\phi$ is plotted against $\xi$ for $M=M_{PPDL}+0.0001$ (solid curve) and $M=M_{PPDL}-0.0001$ (dash-dot curve).}
\end{figure}

\begin{figure}
  \includegraphics{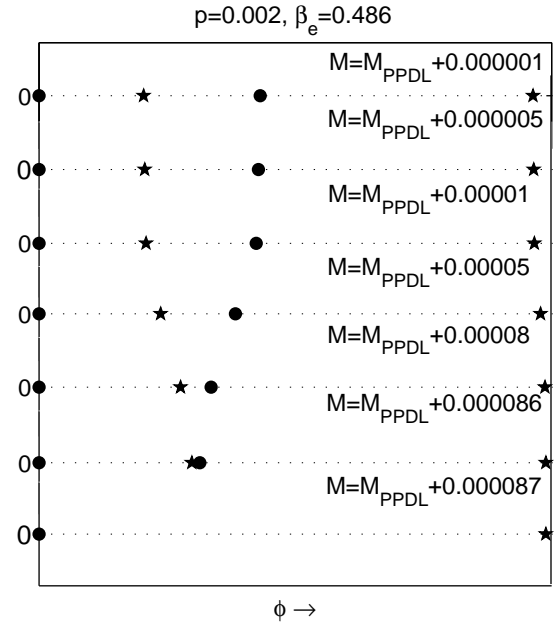}
  \caption{\label{supersoliton_fixed_points} Saddle points (small solid circles) and the equilibrium points other than saddle points (small solid stars) for the system (\ref{phase_portraits}) have been drawn on the $\phi$-axis for the different values of the mach number.}
\end{figure}

From Figure \ref{sol_spc_wrt_beta_e_p=0} - Figure \ref{sol_spc_wrt_beta_e_p=0_pt_02}, we see that for fixed values of $\sigma_{ie}$ and $\sigma_{pe}$ whenever there is no positron in the system, it does not support any PPDL, whereas, the existence of NPDL and coexistence of solitary waves of both polarities are observed for strong nonthermality of electrons. This result is expected, as because in this case, there are no free positive charges. The only positive charges are ions and the concentration of ions is controlled by the concentration of electrons via the Poisson equation and the charge neutrality condition. As the nonthermality of electrons increases, the potential drop of the system optimizes in favour of negative potential. So, the system supports NPDLs for higher value of nonthermal parameter. Now, if we inject isothermal positrons then up to a cut-off value of $p$, say, $p=p^{(a)}$, the qualitative behaviour of the existence domains remain unchanged. But, with the increment of positrons in the system the electrons need stronger nonthermality to optimize the potential drop in favour of negative potential and as a result, the existence region of NPDLs decreases with the increment in $p$. But as $p$ exceeds the cut-off value $p=p^{(a)}$, then up to $p=p^{(b)}$ there exist a region in the parameter space where the saturation level of concentration of total positive charges, concentration of total negative charges and the nonthermality of electrons are such that the system automatically optimizes the potential drop at any point of a particular region of the parameter space in favour of positive potential. Consequently, the system supports PPDLs in that particular region. Again, as $\beta_{e}$ increases and exceeds a critical value $\beta_{e}^{(d)}$, the potential drop optimizes in favour of negative potential and the system starts to support NPDLs. If we increase the concentration of positron from $p=p^{(b)}$, then the strong nonthermality of electrons cannot make the suitable environment to optimize the potential drop towards negative potential, rather, for higher values of $\beta_{e}$, the potential drop optimizes towards positive potential. Therefore, for $p>p^{(b)}$, the system does not support any NPDL whereas PPDLs are observed for higher values of $\beta_{e}$ up to a certain cut-off value of $p$, say, $p=p^{(c)}$. For further increment in the concentration of positron, i.e., for $p>p^{(c)}$, the system does not support any PPDL; there exist only PPSWs which are bounded by the curves $M=M_{c}$ and $M=M_{max}$. Thus not only the concentration of positron but also the nonthermality of electrons plays an important role for the formation of double layers. Beside this, the strong nonthermality of electrons fails to form double layers of any polarity whenever the concentration of positron exceeds a certain cut-off value.

To conclude, we like to mention that the results of the present paper regarding the formation of double layers should be helpful in understanding the possible casues of acceleration of energetic particles in various astrophysical environments, where e-p-i plasma exist. For example, the phenomenon of acceleration of particles in the auroral zone of the atmosphere is due to the double layers which are often generated in the magnetosphere of the Earth \citep{alfven1981}.

We have already mentioned in our earlier paper \citep{paul2016} that till now there is no direct evidence for the existence of supersolitons in both space and laboratory plasma \citep{singh15}. However, further measurements of electric field in space plasma environments by means of satellite expeditions may be able to discover the signature of positive potential supersolitons. We hope that this investigation will add something new ideas regarding the formation of supersolitons to the present knowledge of the nonlinear wave propagation in plasmas.

\acknowledgments The authors are grateful to Professor K. P. Das, Department of Applied Mathematics, University of Calcutta, for valuable discussions. One of the authors (Ashesh Paul) is thankful to the Department of Science and Technology, Govt. of India, INSPIRE Fellowship Scheme for financial support. 


\providecommand{\noopsort}[1]{}\providecommand{\singleletter}[1]{#1}%


\end{document}